\documentclass[a4paper,10pt,twocolumn]{article}
\usepackage{graphicx}
\usepackage{amsmath}
\usepackage[latin1]{inputenc}
\usepackage[english]{babel}

\usepackage{amssymb}
\usepackage{hyperref}
\usepackage[noadjust]{cite} 
\def\01{\lbrace 0,1\rbrace}

\begin{document}

\title{Is there a {fractional} breakdown of the Stokes-Einstein relation in Kinetically Constrained Models at low temperature?}

\author{O. Blondel$^*$ \and C. Toninelli\footnote{\textsc{LPMA, CNRS UMR 7599, Universit\'e Paris VI-VII, B\^atiment Sophie Germain,} {5 rue Thomas Mann, 75205 Paris CEDEX 13, France}
\textit{Email address: }oriane.blondel@ens.fr}}

\maketitle

\begin{abstract}
 We study the motion of a tracer particle injected in facilitated models which are used to model supercooled liquids in the vicinity of the glass transition. We consider the East model, FA1f model and a more general class of non-cooperative models. For East previous works had identified a fractional violation of the Stokes-Einstein relation with a decoupling between diffusion and viscosity of the form $D\sim\tau^{-\xi}$ with $\xi\sim 0.73$. We present rigorous results proving that instead  {$\log(D)=-\log(\tau)+O(\log(1/q))$, which implies at leading order} $\log(D)/\log(\tau)\sim -1$ for very large time-scales. {Our results do not exclude the possibility of SE breakdown, albeit non fractional. Indeed extended numerical simulations by other authors show the occurrence of this violation and our result suggests  $D\tau\sim 1/q^\alpha$, where $q$ is the density of excitations.} For FA1f we prove fractional Stokes Einstein in dimension $1$, and $D\sim\tau^{-1}$ in dimension $2$ and higher, confirming previous works. Our results extend to a larger class of non-cooperative models.
\end{abstract}


\section{Introduction}
A microscopic understanding of the liquid/glass transition and of the glassy state of matter remains a challenge for condensed matter physicists (see \cite{BB,BG} for recent surveys). In the last years many experimental and theoretical works have been devoted to understanding the spatially heterogeneous relaxation which occurs when temperature is lowered towards the glass transition \cite{SwallenEdiger,CE,EEHPW,CS,Swallenetal03,Mapes,Ediger,sillescu,richert,vidal,glotzer,hedges,het}. In this regime dynamics slows down and relaxation is characterized by the occurrence of correlated regions of high and low mobility whose typical size grows when temperature decreases. One of the most striking experimental consequences of dynamical heterogeneities is the violation of Stokes-Einstein relation, namely the decoupling of self-diffusion coefficient ($D$) and viscosity ($\eta$). In high temperature homogeneous liquids, self-diffusion and viscosity are related by the Stokes-Einstein relation $D\eta/T\sim const$ \cite{HM}. Instead in supercooled fragile liquids the self-diffusion coefficient does not decrease as fast as the viscosity increases and $D\eta$ increases by 2-3 orders of magnitude approaching the glass transition \cite{SwallenEdiger,CE,EEHPW,CS,Swallenetal03,Mapes}. A good fit of several experimental data is $D\sim \eta^{-\xi}$ with $\xi<1$ an exponent depending on the specific liquid. Such a violation is instead absent or much weaker in strong liquids, consistently with the idea that the decoupling is related to heterogeneities which are indeed more important for more fragile liquids. A natural explanation of this effect is that different observables probe differently the underlying broad distribution of relaxation times \cite{Ediger}: $D$ is dominated by the more mobile particles, while $\eta$ probes the time scale needed for every particle to move. 

Different theories of the glass transition have been tested by measuring their capability to predict Stokes-Einstein breakdown. In particular, several works \cite{junggarrahanchandler,junggarrahanchandler2,C,BCG} have analysed the self-diffusion coefficient of a probe particle injected in a facilitated (or kinetically constrained) model.

\section{Kinetically constrained models and Stokes-Einstein violation}
In the setting of Kinetically Constrained Models, supercooled liquids are modeled by a coarse-grained mobility field evolving with a Markovian stochastic dynamics with simple thermodynamic properties and non-trivial kinetic constraints. More precisely, facilitated models are lattice models described by configurations $\lbrace n_i\rbrace$, $n_i=0,1$, with $n_i=1$ if the lattice site $i$ is active and $n_i=0$ if $i$ is inactive. Active and inactive sites essentially correspond to coarse grained unjammed and jammed regions, respectively. Active sites are also called defects. The dynamics is described by the following transition rates
\begin{eqnarray}
n_i=0&\overset{qc_i}{\longrightarrow}&n_i=1\\
n_i=1&\overset{pc_i}{\longrightarrow}&n_i=0,
\end{eqnarray}
where $c_i$ encodes the model dependent constraints and is zero or one depending on the local configuration around $i$, $q=1/(1+\exp(1/\tilde{T}))$, $p=1-q$ and $\tilde{T}$ is a reduced temperature. Since $c_i$ does not depend on the configuration on $i$, the dynamics satisfies detailed balance w.r.t. the product measure that gives weight $q$ to active sites and $p$ to inactive sites, which is therefore an equilibrium distribution. Two very popular models are the one-spin facilitated model, FA1f \cite{FA1}, and the East model \cite{eastphys}. For FA1f, $c_i=1$ iff site $i$ has at least an active nearest neighbour, while for East in one dimension $c_i=1$ iff the right neighbour of $i$ is active (namely $c_i=n_{i+1}$). The injection of a probe particle into these models is performed as follows \cite{junggarrahanchandler,junggarrahanchandler2}. Initially the lattice configuration is distributed with the equilibrium product measure and the probe particle is at the origin. Then one lets the lattice configuration (the environment) evolve according to the facilitated model dynamics while the probe is allowed to jump only between active sites, namely 
\begin{equation}
X {\longrightarrow} X\pm e_\alpha\quad\text{ at rate }\quad{n_Xn_{X\pm e_\alpha}}
\end{equation}
where $X$ is the position of the probe, $\alpha=1,...,d$ is one of the $d$ directions and $e_\alpha$ is the unit vector in this direction. Then the self-diffusion matrix $D$ is defined as usual by $$e_\alpha.2De_\alpha=\lim_{t\rightarrow\infty}\frac{\left\langle \left(X_t\cdot e_\alpha\right)^2\right\rangle}{t}.$$

A numerical analysis for the FA1f model leads in \cite{junggarrahanchandler,junggarrahanchandler2} to the conclusion that $D\sim q^2$ in any dimension. Previous numerical \cite{BG} and renormalisation group analysis \cite{WBG} suggested $\tau=1/q^{2+\epsilon(d)}$ with $\epsilon(1)=1$, $\epsilon(2)\simeq 0.3$, $\epsilon(3)\simeq 0.1$ and $\epsilon(d\geq 4)\simeq 0$. These estimates led \cite{junggarrahanchandler,junggarrahanchandler2} to the conclusion that Stokes-Einstein relation is violated with $\xi\simeq 2/3,2/2.3,2/2.1$ for FA1f in $d=1,2,3$ and is not violated in higher dimensions. In \cite{goodtau} the scaling of $\tau$ was deduced via an exact mapping into  a model of annihilating random walks with spontaneous creation from the vacuum $A+A\leftrightarrow 0$, leading instead to $\epsilon(d\geq 2)=0$. This finding is supported by the mathematical results in \cite{CMRT} which confirm $\epsilon(2)=0$ and yield $\epsilon(3)\leq 0$. In consequence the result for the diffusion coefficient in \cite{junggarrahanchandler,junggarrahanchandler2} was reinterpreted \cite{garrahansollichtoninelli} by saying that $\xi=2/3$ in $d=1$ while no violation occurs in $d\geq 2$. This is consistent with the idea that FA1f is a non cooperative model dominated by the diffusion of active sites and it is a model for strong rather than for fragile liquids. Instead for the East model the analysis in \cite{junggarrahanchandler,junggarrahanchandler2} leads to $D=\tau^{-\xi}$ with $\xi\simeq 0.73$, a result which is expected to hold also in higher dimensions. The exponent is consistent with the one observed experimentally and numerically in fragile glass-forming liquids \cite{Swallenetal03},\cite{yamamotoonuki},\cite{berthier04}.

Here we report recent rigorous mathematical results for the East and FA1f models and for more general non-cooperative models (details can be found in \cite{preprint}). For the one dimensional East model we prove that there exist constants $\alpha$,{$c_1,c_2>0$} such that
{\begin{equation}\label{eastres}
c_1q^2\tau^{-1}\leq D \leq c_2q^{-\alpha}\tau^{-1}
\end{equation}}
which yields at leading order
{\begin{equation} \label{eastresleading}
\frac{\log(D)}{\log(\tau)}=-1+o(1),
\end{equation}
where $o(1)$ is a term that vanishes at low temperature}, since $\tau$ diverges faster than polynomial as $q\to 0$. Thus we establish that a fractional Stokes-Einstein relation cannot hold, in contrast with the predictions in \cite{junggarrahanchandler,junggarrahanchandler2}. 
The numerical results in the latter works clearly show that a SE violation occurs. Our result \eqref{eastres} does not exclude the occurrence of a SE violation, albeit non fractional. 
In particular a (weaker than fractional) violation compatible with our result and with our heuristic
is $D\tau\sim 1/q^{\alpha}$.
We provide a heuristic for our result, which is related to the estimate of the energy barriers that the probe has to overcome in order to cross the typical distance between two active sites at equilibrium. {In fact, the probe typically can do no better to exploit the underlying fluctuation of the East model than jump a distance $1/q$ in time $\tau$.} We also provide our understanding of which are the problems in the analysis performed in previous works. Then we consider non-cooperative models and we prove, {in agreement with \cite{junggarrahanchandler}}, that in any dimension for FA1f it holds
\begin{equation}\label{resFA1f}
cq^2\leq D\leq c' q^2,\end{equation}
with $c,c'$ constants independent on $q$. We also prove
\begin{equation}\label{reskdef}
cq^{k+1}\leq D\leq c' q^{k+1}\end{equation} 
for a more general model in which $k$ (instead of one) active sites are required in the vicinity of the to-be-updated site. We provide a heuristic both for the diffusion coefficient and the relaxation time which leads to a fractional Stokes-Einstein for $d=1$ and to $D\sim\tau^{-1}$ for $d\geq 2$. In particular our heuristics clearly explain the scaling $\tau=1/q^2$ in $d\geq 2$ for the FA1f model. Note that \eqref{resFA1f} together with the results in \cite{CMRT} imply that for FA1f in $d\geq 3$ it holds $D\tau\leq const$: any form of decoupling cannot hold in this case (while a logarithmic decoupling may occur in $d=2$). Finally we obtain for any choice of the kinetic constraints a variational formula for the diffusion matrix, which we will present and discuss at the end in order to avoid technicalities at this stage. As a consequence we obtain for any facilitated model
\begin{equation}\label{eq:bornesfaciles}
q^2\tau^{-1}\leq e_{\alpha}.De_{\alpha}\leq q^2.
\end{equation}

\section{East model}

The relaxation time of the East model has an exponential inverse temperature squared (EITS) form. Namely, up to polynomial corrections,
\begin{equation}\label{eq:taueast}
\tau\sim e^{\ln(1/q)^2/2\ln 2}.
\end{equation} 
The form $\tau\sim e^{cst/ T^2}$ was first given in \cite{taueastphys} with $cst=1/\ln 2$, which was derived via energy barrier considerations. This value of the constant was proved to be wrong by a factor $1/2$ in \cite{KCSM}. Indeed, taking into account an entropy factor which was missing in the previous works (see also \cite{eastrecent} for a more extended explanation) and using the lower bound of \cite{aldousdiaconis}, in \cite{KCSM} it was proven instead that $cst=1/2\ln2$. This scaling can be explained through combinatorics arguments. Consider a configuration of only inactive sites on a typical equilibrium length $1/q$, with a fixed active site at the right boundary. Recall that, due to the orientation of the constraint, the left-most site can only become active if all sites on its right became active before it. It was proven in \cite{taueastphys,combinatoricseast} that before the leftmost site can become active, the system needs to visit configurations with at least $\ln(1/q)/\ln 2$ active sites. The equilibrium probability of such a configuration is less than $e^{-\ln(1/q)^2/\ln 2}$ when $q\rightarrow 0$, which accounts for the EITS form. Moreover, the set of configurations attainable using at most $n=\ln(1/q)/\ln 2$ active sites simultaneously has a cardinality of order $2^{\binom{n}{2}}n!\approx e^{\ln(1/q)^2/2\ln 2}$ \cite{combinatoricseast}, so that the entropy factor changes the constant in the EITS form by a factor $2$ and yields \eqref{eq:taueast}. This fast divergence of $\tau$ makes it very difficult to approach zero temperature through simulations and allows to neglect polynomial terms in $q$ when an estimate involves $\tau$. The above discussion actually explains the scale of the \emph{persistence} time rather than the relaxation time. However, for the East model these characteristic times coincide \cite{timescalesepphys}. Let us provide the heuristics behind our result \eqref{eastres} which establishes that also diffusion occurs on this time scale at leading order. In the initial configuration, the first active site ($i_a$) on the right of the probe particle is typically at distance $\sim 1/q$. Before the tracer can move its first step to the right it needs at least to wait for its right neighbour to become active. This occurs thanks to the fact that sites are activated from right to left starting from $i_a$ and thus requires a time proportional to the persistence time. Note that the arrival of the excitation sent from $i_a$ does not influence the configuration on the right of $i_a$. In particular once the probe has arrived at $i_a$ it has typically to face again the same energy barrier. In summary, for each distance of $1/q$ the probe covers towards the right we need a time at least $\tau$ and this, together with the symmetry of the motion of the probe and the fact that any polynomial in $q$ is negligible with respect to $\tau$, yields \eqref{eastresleading}. Note that our result \eqref{eastres} allows a weaker violation of the Stokes Einstein relation: $D\tau$ can diverge when $q\to 0$ as a polynomial in $1/q$ {and our heuristics suggest that the power of this polynomial should be at most two.} Indeed, recent and more extended simulations \cite{comment} are compatible with $D\tau\sim 1/q^\alpha$ with $\alpha\sim 1.6$.

We believe that the discrepancy between our result and the findings $D\sim \tau^{-\xi}$ with $\xi\sim 0.73$ in \cite{junggarrahanchandler,junggarrahanchandler2} is due the difficulty to approach zero temperature in simulations. In particular, among the diffusion coefficient data reported on Fig.3 of \cite{junggarrahanchandler}, on all data except the last one the value of $1/T$ is such that $1/q^2 >e^{\ln(1/q)^2/2\ln 2} $. Thus these data, even though very accurate and asymptotic in time, are not sufficiently in the low temperature regime and do not allow to capture the asymptotic form of $D$ vs $\tau^{-1}$ when $q\to 0$.\footnote{After extended and fruitful discussions, the authors of \cite{junggarrahanchandler,junggarrahanchandler2}, performed new and much more extended numerical simulations \cite{comment}. In the numerically accessible range, their data are still compatible with the fractional violation which is excluded by our asymptotic result. In view of our findings, a new fit with a weaker polynomial violation $D\tau\simeq 1/q^\alpha$ was performed. This form is also compatible with the numerical data. This confirms that in the low density regime the analysis of numerical simulations is very delicate due to the extremely slow dynamics.} The presumed fractional decoupling for East was considered (see e.g. \cite{BG},\cite{ChGa}) to be a consequence of the fluctuations in the dynamic. More precisely it was explained by the fact that, even if the first move is governed by the persistence time, then the probe is supposed to move faster since the typical time for the next events was considered to be the (shorter) mean time between changes of mobility for a given site (exchange time). To use the expression of \cite{ChGa}, the probe should surf on excitation lines and thus move faster than the typical relaxation time. Due to the directed nature of the constraint, the excitation line cannot expand to the right of the site where it has originated, therefore the probe can perform this fast surfing only up to a distance $1/q$: the persistence time remains the leading order in the diffusion time scale while fluctuations should give rise to a polynomial violation of Stokes-Einstein.

\section{FA1f and other non cooperative models}

We turn now to non-cooperative models, and more specifically to the $k$-defects model which we define as follows: $c_i=1$ if and only if there are at least $k$ defects at distance at most $k$ around $i$. Note that for $k=1$, we recover the FA1f model and that any $k$-defects model is non-cooperative: if the initial system contains $k$ active neighbours, any site can be activated through allowed transitions. Also, at low $q$ we expect dynamics to be dominated by the diffusion of the group of $k$ defects, which occurs at rate $q$ because in order to shift of one step the group of vacancies we need to create an additional vacancy in the direction of the move (and then remove one vacancy of the group in the opposite direction). As stated in \eqref{reskdef}, we prove in all dimensions $D\sim q^{k+1}$, which agrees with the numerical results in \cite{junggarrahanchandler} for FA1f ($k=1$). The heuristics behind \eqref{reskdef} is the following. Consider a box of size $q^{-k}$ centred on the probe particle. Typically at equilibrium there is one group of $k$ active sites inside this box, so that the proportion of time during which the probe particle is on such a group is $q^k$. During that portion of time, the probe particle diffuses at the same rate as this group of $k$ active sites which, as already explained, is $q$. In the end, the diffusion coefficient of the probe particle is of order $q^k\times q=q^{k+1}$. Concerning the relaxation time we expect $\tau\sim 1/q^{2k+1}$ in one dimension and $\tau\sim 1/q^{k+1}$ in $d\geq 2$. This, together with \eqref{reskdef}, implies that a fractional violation of the Stokes Einstein relation does not occur in $d\geq 2$ and occurs in $d=1$. In $d=1$ the result for $\tau$ should come from the fact that relaxation requires the group of $k$-vacancies to overcome the typical distance $1/q^k$ among two subsequent groups by diffusing at rate $q$. In $d\geq 2$ around each group of $k$-defects there is typically a ball of radius $r=1/q^{k/d}$ without any such group. Relaxation requires that a fraction of the sites of the ball is covered by the active group which is essentially a random walker with rate $q$. Classic results on random walks \cite{aldous,dembo} imply that this requires a time (up to log corrections) $r^d$ times the inverse of the diffusion rate of the walker, which indeed yields $\tau\sim 1/q^{k+1}$.

Before sketching the ideas that allow us to prove \eqref{reskdef} rigorously, we wish to present our variational formula
for the diffusion matrix, which is valid for any choice of the constraints and in particular yields \eqref{eq:bornesfaciles}.
Denote by $\eta_i(t)$ the state of site $X_t+i$ at time $t$, i.e. $\eta(t)$ is the configuration \emph{seen from the probe particle} at time $t$. In particular, the state of the system at the position of the tracer at time $t$ is given by $\eta_0(t)$. We call $j_\alpha$ the current of the probe in the direction $\alpha=1,...,d$, namely
\begin{equation}
j_\alpha(\eta)=\eta_0\left(\eta_{e_{\alpha}}-\eta_{-e_{\alpha}}\right).
\end{equation}
Finally, we denote by $\mathcal{L}$ the Liouvillian operator associated to the master equation for the dynamics i.e. $\mathcal{L} $ is the operator such that $\partial _t \langle f(\eta(t))\rangle=-\langle \mathcal{L}f(\eta(t))\rangle$, where $\langle \cdot\rangle$ denotes the mean over trajectories and over the initial configuration distributed with the equilibrium measure. This is the adjoint of the operator $\mathbb{W}$ governing the master equation: $\partial_t|P\rangle =-\mathbb{W}|P\rangle$. We use this operator to express the typical value of $f$ at time $t$ as $\langle f(\eta(t))\rangle =\langle e^{-\mathcal{L}t}f\rangle$. Note that $\mathcal{L}=\mathcal{L}_{env}+\mathcal{L}_{jump}$, where $\mathcal{L}_{env}$ is the Liouvillian operator for the evolution of the environment (the facilitated model without the probe), and $\mathcal{L}_{jump}$ describes the evolution caused by the jumps of the probe particle. Using standard methods \cite{spohn} we compute the limit of the rescaled position of the probe particle in terms of the current and get the following result for $e_\alpha.2De_\alpha$ \cite{preprint}
\begin{eqnarray}
\sum_{y=\pm e_\alpha}\left\langle\eta_0\eta_{\pm e_\beta}\right\rangle - \lim_{t\rightarrow\infty}\frac{1}{t}\left\langle\left(\int_0^t j_\alpha(\eta(s))ds\right)^2\right\rangle,\label{eq:limit}
\end{eqnarray}
where $\langle \cdot\rangle$ has the same meaning as above. In \eqref{eq:limit}, the first term is just $2q^2$ and the second one is $-\int_0^\infty\left\langle j_\alpha(\eta(0))j_\alpha(\eta(s))\right\rangle$, which is $-2\int_0^\infty\left\langle j_\alpha e^{-t\mathcal{L}}j_\alpha\right\rangle$ in the above formulation and can be rewritten as $2\left\langle j_\alpha\mathcal{L}^{-1}j_\alpha\right\rangle=-2\inf_f\left\{2\mu(j_\alpha f)-\left\langle f\mathcal{L}f\right\rangle\right\}$. Then some computations (see \cite{preprint} for details) yield the following variational formula for $e_\alpha.2De_\alpha$:
\begin{equation}\label{eq:varform}
\inf_f\left\{2\left\langle f\mathcal{L}_{env}f\right\rangle+\sum_{y=\pm e_\alpha}\left\langle\eta_0\eta_y\left[y_\alpha+f(\tau_{y}\eta)-f(\eta)\right]^2\right\rangle\right\},
\end{equation}
where $\langle\cdot\rangle$ denotes the mean w.r.t. the equilibrium measure and $\tau_y\eta$ is $\eta$ translated by the vector $y$.

\begin{figure}
\begin{center}
\includegraphics[scale=0.2]{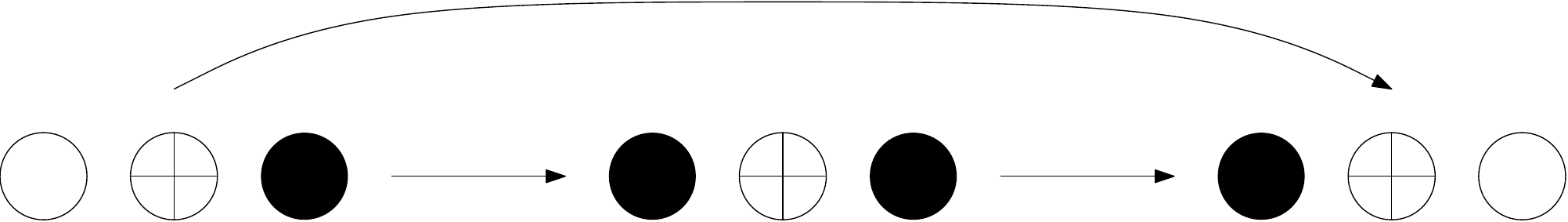}
\caption{{Active (inactive) sites are in white (black) and the probe is marked by a cross. On the left, a configuration at equilibrium with the probe at the origin, an active site at the origin and an active site on the left of the probe. On the right, the swapped configuration (the swap materialized by the upper arrow is the only transition allowed in the auxiliary dynamics, apart from the jumps of the probe). It is possible to reconstruct this swap (\emph{i.e.} go from the configuration on the left to the one on the right), using only flips allowed by FA1f and without adding extra active sites.
Indeed one can start by creating an inactive site on the active neighbour
of the probe (middle configuration)
and then reach the final configuration by creating an active site on the
other neighbour (both moves are allowed by FA1f rates thanks to the active
site on which the probe sits).}}
\label{fig:swapdoneinflips}
\end{center}
\end{figure}

We are now ready to sketch the ideas that allow us to prove \eqref{reskdef}. To establish $D\geq cq^{k+1}$, we show that $D\geq cq^{k+1}\overline{D}$, where $\overline{D}$ is the diffusion coefficient of a $k$-dependent auxiliary dynamics which we describe in the case $k=1$ (FA1f) in dimension one. Take an initial configuration at equilibrium, with the probe at the origin, an active site at the origin and at least an active site among its neighbours. Then define the auxiliary dynamics as follows. The probe particle can jump to a neighbouring active site with rate $1$, and the two neighbours of the probe particle can swap: if one of them is active and the other inactive, they exchange their activity state with rate $1$. Note that with these rules the probe particle is always on an active site and has always an active neighbour. In particular, we can show that the diffusion coefficient for this auxiliary dynamics $\overline{D}$ is positive and does not depend on $q$ {(see also \cite{spohn}). Then we need to establish $D\geq cq^{2}\overline{D}$ to conclude. This is possible because we can compare the formula \eqref{eq:varform} with its analogue for $\overline{D}$, the diffusion coefficient in the auxiliary dynamics. In fact, the crucial ingredient is that it is possible to reconstruct any possible move in the auxiliary dynamics using a finite number of moves allowed by the FA1f dynamics (see Fig.~\ref{fig:swapdoneinflips}). As a consequence, the first term in \eqref{eq:varform} can be compared with the analogous in the variational formula for the diffusion coefficient of the auxiliary dynamics. The important thing in this reconstruction is that intermediate steps involve no extra active site and therefore no extra factor $q$ comes out of this comparison.} The term $q^2$ comes from the cost of imposing an active site at the origin and on one of its neighbours in the equilibrium configuration. The extension to other values of $k$ and higher dimensions are detailed in \cite{preprint}.

\begin{figure}
\begin{center}
{\includegraphics[scale=0.6]{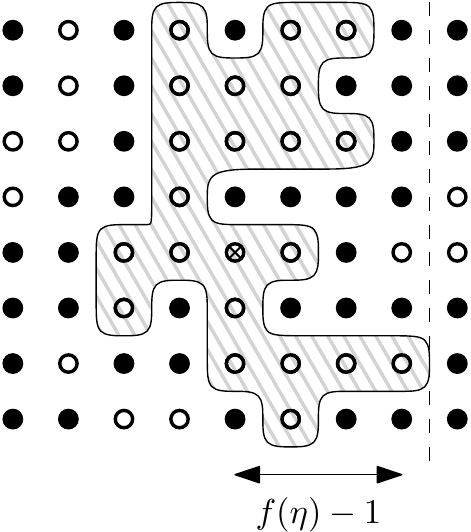}}
\caption{{Here the origin is crossed. The connected cluster of active sites we consider to define our test function $f$ is hatched, its maximal extension to the right is marked by a dashed line. We read $f(\eta)=4$.}}
\label{fig:fonctiontest}
\end{center}
\end{figure}

In order to show $D\leq C q^{k+1}$, we look for an observable $f$ that captures the order of the diffusion when plugged in the variational formula \eqref{eq:varform}. We treat the case $\alpha=1$. In a configuration at equilibrium, consider the connected cluster of active sites containing the origin. This is the cluster that the probe could span if the environment remained frozen (see Fig.~\ref{fig:fonctiontest}). {Given a configuration $\eta$,} we choose $f(\eta)$ to be the smallest non-negative coordinate $z$ such that this cluster is contained in the half-space on the left of $z$, and we let $f(\eta)=0$ if the origin is inactive (see Fig.~\ref{fig:fonctiontest} for an example). The calculations in \cite{preprint} show that the test function $f$ captures indeed the correct behaviour of the diffusion matrix.

\section{Conclusion}

In summary, we proved that for the East model in dimension one the self-diffusion coefficient of a probe particle is such that {$\log(D)/\log(\tau)\sim -1$} in the low temperature regime ($q\rightarrow 0$), at variance with previous results claiming a fractional Stokes-Einstein relation of the form $D\sim\tau^{-\xi}$ with $\xi<1$. Our results suggest a weaker violation of the form $D\tau\sim1/q^\alpha$. We also establish a variational formula for $D$ which is valid for any kinetically constrained spin model in the ergodic regime. For FA1f model and more generally ``$k$-defects" models, a detailed study of this variational formula allowed us to prove the exact order of the diffusion coefficient: $D\sim q^{k+1}$. This, together with the heuristics we provide for the scaling of the relaxation time, implies a fractional breakdown of the Stokes-Einstein relation only in dimension one. 
 
In \cite{junggarrahanchandler2} higher dimensional generalisations of the East model have been considered and a fractional Stokes-Einstein with $\xi\sim 0.7-0.8$ weakly dimensionally dependent has been observed. {Since the relaxation time is again larger than any polynomial in $1/q$ and the distance of the active sites is $1/q^{1/d}$, again a decoupling cannot occur as a consequence of the difference between persistence and exchange times and we expect no fractional violation either. Recent rigorous results \cite{east_Zd} moreover show that persistence and relaxation times are of the same order in infinite volume dynamics. However, the authors also evidence highly non-trivial behaviour of these characteristic times in finite volume (in particular an anisotropy phenomenon); extending our mathematical proof to higher dimensions would require a deep understanding of the subtle energy-entropy competition studied in \cite{east_Zd}.}

In the future, we also wish to investigate
other cooperative models such as Fredrickson-Andersen two spin facilitated model (FA2f) \cite{FA1} or the spiral model \cite{spiral}. In this case the event which triggers the moves of the probe could be more cooperative and it could modify the configuration up to a distance larger than a polynomial in $1/q$. Thus the fractional violation of Stokes Einstein observed in supercooled liquids could be reproduced by these kinetically constrained models.

\section{Acknowledgements}
We acknowledge very useful discussions with G.Biroli and T.Bodineau and thank V. Lecomte for comments on a draft version.

\bibliography{bibliotracer}
\end{document}